\documentclass[sigplan,nonacm]{acmart}
\setkeys{acmart.cls}{balance=false}
\usepackage{amsmath}
\usepackage{graphicx}
\usepackage{textcomp}
\usepackage{xcolor}
\usepackage{xspace}
\usepackage{tikz}
\usepackage{comment}
\usepackage[font=footnotesize,labelfont=bf]{caption}
\usetikzlibrary{positioning, calc, fit, arrows.meta, shapes.geometric, backgrounds}
\usepackage{enumitem}
\usepackage{array}
\usepackage{xurl}
\AtBeginEnvironment{thebibliography}{\raggedright}

\setlist[itemize]{topsep=2pt, partopsep=0pt, itemsep=2pt,
                  parsep=0pt, leftmargin=1.2em}

\definecolor{dong}{RGB}{0,0,200}
\definecolor{bin}{RGB}{200, 100, 50}

\newcommand{\name}{CARDAN\xspace}

\begin{document}
\raggedbottom

\title{A Multi-Engine Dataflow for MoE Decoding on AWS Tranium}
\renewcommand{\shorttitle}{}
\renewcommand{\shortauthors}{}

\author{Bin Ma}
\affiliation{
  \institution{University of California, Merced}
  \city{Merced}
  \state{California}
  \country{USA}
}
\email{bma100@ucmerced.edu}

\author{Wenjie Fan}
\affiliation{
  \institution{Yotta Labs}
  \country{USA}
}
\email{fanwj@mail.ustc.edu.cn}

\author{Jonny Liu}
\affiliation{
  \institution{Yotta Labs}
  \country{USA}
}
\email{liu@yottalabs.ai}

\author{Dong Li}
\affiliation{
  \institution{University of California, Merced}
  \institution{Yotta Labs}
  \city{Merced}
  \state{California}
  \country{USA}
}

\email{dli35@ucmerced.edu}

\begin{abstract}
Mixture-of-Experts (MoE) decoding on scratchpad-based tensor accelerators (STA) is dominated by moving expert weights while the compute engines sit idle. This traffic is hard to hide, because the experts are known only after routing, and hard to shrink without losing quality or adding critical-path work. We present \name, which represents each expert-weight matrix as a vector-quantized component plus a shared-basis low-rank component and co-designs this representation with a multi-engine decoding dataflow.
The representation separates expert-common from expert-private work, so the dataflow overlaps DMA with computation on several engines. Across five MoE families on AWS Trainium3, CARDAN matches or improves BF16-teacher perplexity across all five models and speeds up batch-one decoding by $1.15$--$1.31\times$ over AWS dense MoE megakernels, rising to $1.7\times$ at batch size 16.

\end{abstract}

\maketitle

\section{Introduction}
\label{sec:intro}

Mixture-of-Experts (MoE) scales language models by activating only a subset of a layer's experts for each token, increasing parameter capacity while keeping per-token floating-point operations (FLOPs) nearly constant. MoE has therefore become a common architecture in recent frontier models.  


The scratchpad-based tensor accelerator (STA) is a specialized hardware system designed for deep learning and matrix workloads. STA relies on a software-controlled Scratchpad Memory (SPM) rather than a traditional hardware managed cache: software places each tensor explicitly on SPM, and Direct Memory Access (DMA) engines move data between high-bandwidth memory (HBM) and SPM. By using explicitly managed on-chip storage, STA eliminates cache tag overhead and unpredictable replacement policies, maximizing energy efficiency and throughput for large-scale mathematical computations.
STA has been exemplified by a range of  accelerators, such as AWS Trainium~\cite{aws2026trainium3arch}, Google TPU~\cite{jax2026tpupipelining}, Intel Gaudi\,2~\cite{intel2022gaudi2}, and Huawei Ascend~\cite{huawei2026ascendcarchitecture}.  
Hence, we study MoE on STA. 

\begin{figure}[!t]
  \centering
  \includegraphics[width=\columnwidth]{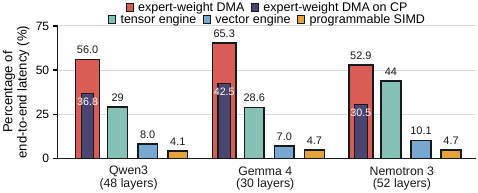}
  \vspace{-18pt}
  \caption{The breakdown of decoding latency for three MoE models from Hugging Face \cite{qwen2025qwen3checkpoint,google2026gemma4checkpoint,nvidia2025nemotron3checkpoint}. The decoding generates one token based on four-way tensor parallelism. ``CP'' stands for critical path.}
  \vspace{-14pt}
  \Description{Grouped bars show the share of token latency occupied by expert-weight transfers and the tensor, vector, and programmable SIMD engines for Qwen3, Gemma 4, and Nemotron 3. Additional bars identify expert-weight transfer time on the critical path.}
  \label{fig:baseline-engine-activity}
\end{figure}

\textbf{Problems.} 
Despite a sparse routing mechanism that activates only a subset of experts per token, MoE decoding latency on STAs is still often dominated by expert-weight DMA.
For example,  Qwen3-30B-A3B has 48 layers, and each layer routes a token to activate only eight of 128 experts, 
generating $75$\,MB of weight traffic (using BF16) per layer per token. The acting expert set may change from token to token, forcing the selected expert weights to be loaded from HBM with little reuse. We run this Qwen3 decoding with four-way tensor parallelism on an example STA, AWS Trainium3 using four logical NeuronCores, each with 64 MiB of SPM. 
Figure~\ref{fig:baseline-engine-activity} shows the results. For Qwen3, 
Loading expert weights takes 56\% of the end-to-end latency, and the loading time is largely (36.8\%) exposed to the critical path.
We have similar observations on Gemma 4 and Nemotron 3: the loading time exposed to the critical path takes a large portion (42.5\% and 30.5\%) of the end-to-end latency.

Addressing the long latency problem due to expert-weight movement faces two challenges.
First, expert-weight traffic is difficult to hide. Prefetch, by overlapping with compute, is an effective approach to hide the data movement overhead, but the experts to prefetch is only known after the router selects experts, which brings limited opportunities for overlap. The recent efforts~\cite{hwang2024pregated,du2024sida,xue2024moeinfinity,song2024promoe} anticipate the routing selection by utilizing history of expert activations. However, those efforts cannot work well on STA, because staging expert weights before routing creates a long-lived buffer in limited SPM, and disrupts the effectiveness of compiler optimization to maximize data locality. Even worse, the compiler optimization can reclaim and overwrite the memory of prefetched experts in SPM \cite{aws2026nkidirectalloc}, invalidating prefetch and wasting data movement.  

Second, reducing the traffic size to reduce the data movement overhead is challenging. Changing the expert-weight representation (e.g., using sparse tensor formats \cite{chen2019eyerissv2,qin2021formats}) can reduce traffic size, but easily disrupts the effectiveness of DMA. The DMA service time depends not only on traffic size, but also on the number of DMA packets, memory address regularity, and transfer serialization \cite{saidi2012dma,benz2024dma}. A smaller weight-representation may replace regular transfers with more finer-grained or irregular ones, and transferring fewer bytes does not proportionally translate to shorter DMA-service time. Existing efforts~\cite{chen2019eyerissv2,qin2021formats} also show that compressed tensor formats can introduce metadata-dependent memory accesses and that the format with the smallest storage footprint can incur longer execution latency. 
Moreover, some tensor representations trade traffic reduction with additional runtime work. For example, vector quantization (VQ) ~\cite{jegou2011pq,liu2024vptq} reduces weight traffic but requires that each expert stores an index map into a code book.  
Selecting the code word in the code book introduces runtime overhead, taking up to 20\% of the end-to-end decoding latency in our evaluation on Trainium.

Besides the data movement problem, we also see under-utilized compute engines when serving MoE decoding on STA. The STA is often characterized with multiple compute engines, such as tensor engine and SIMD engine, each of which is good at handling workloads with specific properties (e.g., memory access patterns and tensor shapes). Take Qwen3 shown in Figure~\ref{fig:baseline-engine-activity} as an example again. The SIMD engine is active only during 4.1\%-4.7\% of serving time, and the vector engine is active only during 7\%-10.1\%. 

However, decomposing the MoE workload and map it to those engines is challenging, because routing, expert-weight transfer, and expert computation form a strict dependency chain, leaving little concurrency to tap without losing MoE quality. Furthermore, decomposing the expert computation (including expensive up- and down-projection) to finer-grained computation tasks and mapping them across engines cannot bring latency reduction because of operand dependencies across engines and extra SPM traffic.

\textbf{Key insights.} We reveal that designing an expert-weight representation for STAs is key to addressing the above problems of data movement and under-utilized hardware. Jointly designing the weight representation and execution dataflow reduces weight transfers from HBM to SPM. It also allows multiple compute engines to work concurrently with DMA transfers.

\textbf{Solutions.} Based on the above insights, we introduce \name, an MoE framework with an expert-weight representation designed for STAs and execution across their compute engines. \name has multiple innovations.

First, \name uses a ``hybrid'' expert-weight representation. To reduce the data movement, we employ the low rank (LR) factorization method~\cite{molae2025,chen2025mobe}. This method was originally used to reduce the memory consumption of MoE parameters by extracting a common basis-tensor per layer from all experts in the same layer; 
besides the basis tensor, each activated expert uses small, private coefficients to reconstruct the expert. To reduce the data movement, at each layer, we put the basis tensor in SPM and \textit{share} it across experts, but sparsely load the small expert-private coefficients. 

However, we find that the LR leads to severe quality loss, because 
the expert-private coefficients only provide low-rank approximation of expert weights and miss critical expert-specific information. 
Using the LR, for example, 
Qwen3-30B-A3B loses the quality: the perplexity, which is the output quality metric and whose lower is better, increases from 9.82 to 34.91. 
Hence, we face a tradeoff between reducing data movement and maintaining MoE quality.


To avoid the quality loss, we use a VQ technique~\cite{linde1980vq} in combination with the LR. The VQ fits STA naturally, because the VQ is based up on a common and small code book shared across experts in a layer. As a result, the code book can reside in SPM, loaded once but reused often, which avoids data movement. Using the VQ, the experts are re-constructed by selectively loading code word from the code book, restoring some expert information missed in the low rank. Meanwhile, the VQ benefits from the LR for reducing the runtime overhead, because the combination of the VQ and LR provides sufficient quality, which allows the VQ to use coarse-grained code words to reduce runtime overhead. As a result, the hybrid LR/VQ brings three-fold benefits: less data movement, low runtime overhead, and high quality. Also, since the basis tensor and code book are small, the compiler has much space in SPM to tap and improve tensor locality, and there is no conflict for performance optimization.

Second, \name leverages the new expert-weight representation to enable higher utilization  of multiple compute engines. Based upon the hybrid expert-weight representation, we are able to break the traditional routing--transfer-compute dependency chain to two parallel phases: one mainly focusing on expert-common computation without the dependency on the routing decision and the other focusing on expert-private computation. Each parallel phase has multiple independent compute and IO tasks, bringing parallelism and overlap of compute/IO. Also, the tasks in each phase have different characteristics and hence enable efficient utilization of different engines.

We deploy \name{} across Qwen3, DeepSeek-V2-Lite, Nemotron, OLMoE and Gemma4 on AWS Trainium3 (an example of STA). Across these deployments, \name accelerates decoding by $1.15$--$1.31\times$ over the state-of-the-art baseline of AWS BF16 dense megakernel, while achieving similar quality. 
Our work spans computer architecture and \textcolor{dong}operating system (generally defined), with contributions to heterogeneous architectures and dataflow runtime system.   
We summarize the major contributions as follows.

\begin{itemize}
    \item We analyze the MoE performance on STA, and identify the data movement as the most significant challenge of deploying MoE on promising STA architectures.

    \item We create a new method to reduce and hide the data movement overhead on STA based on a hybrid expert-weight representation. This method breaks the traditional data-dependency barrier for parallelism, and allows us to tap heterogeneous compute engines in STA.

    \item We build a framework \name based on the above method and demonstrate the feasibility of our method on a representative STA, AWS Trainium3.
\end{itemize}

\section{Background}
\label{sec:bg}
We review the background information for \name. Table~\ref{tab:notation} summarizes the major notations used throughout the paper.

\subsection{MoE Background}
\label{sec:bg-moe}

\begin{figure}[t]
  \centering
  \includegraphics[width=\columnwidth]{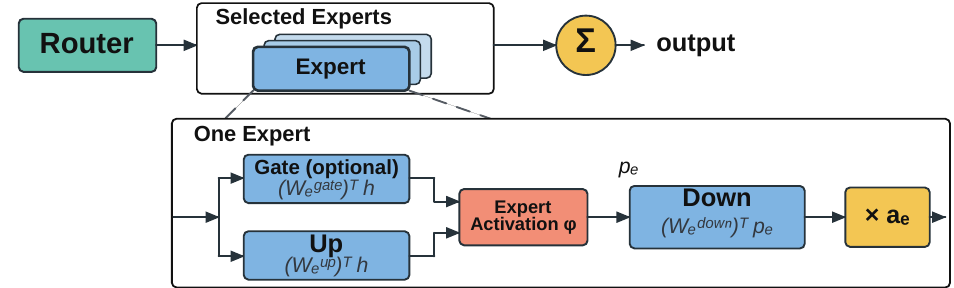}
  \caption{Routing and expert computation in MoE.}
  \Description{The router selects experts for a token. Each selected expert applies gate and up projections, an activation, and a down projection. Routing affinities weight the expert outputs before their sum.}
  \label{fig:moe-computation}
\end{figure}

An MoE layer has an attention block followed by an MoE block. The MoE block contains a router and a pool of experts.
Figure~\ref{fig:moe-computation} depicts the MoE block. Given an input activation $h\in\mathbb{R}^{H}$, the router selects a set of $k$ experts and assigns a expert-specific routing affinity $a_e$ to each selected expert. 
Each selected expert applies its gate and up projection to $h$, combines their results 
through the model-specific expert activation $\phi$
to produce the expert intermediate activation $p_e\in\mathbb{R}^{I}$, and maps $p_e$ back to the hidden dimension through the down projection. The MoE block scales each selected expert's output by its $a_e$ and sums the scaled outputs. 
We define each expert's gate, up projection, and down projection matrices as \emph{expert-weight matrices}, denoted generally by $W_e$. The \textit{expert computation} comprises the gate, the two projections, and the expert activation.




\begin{table}[t]
\centering
\caption{Notations.}
\label{tab:notation}
\scriptsize
\renewcommand{\arraystretch}{0.90}
\setlength{\tabcolsep}{1.5pt}
\begin{tabular}{@{}p{0.14\columnwidth}@{\hspace{2pt}}>{\raggedright\arraybackslash}p{0.3\columnwidth}@{\hspace{8pt}}p{0.14\columnwidth}@{\hspace{2pt}}>{\raggedright\arraybackslash}p{0.3\columnwidth}@{}}
\hline
symbol & meaning & symbol & meaning \\
\hline
$h,H$ & Input activation; hidden dimension. & $p_e,I$ & Expert intermediate; dimension. \\
$e,E,k$ & Index; total/selected expert counts. & $a_e$ & Routing affinity. \\
$W_e,\widehat W_e$ & Expert matrix; hybrid approximation. & $d,K,r$ & VQ block length; weight code book size; rank. \\
$\mathcal B$ & Shared VQ weight code book. & $J_e$ & Private VQ index map. \\
$D$ & Shared low-rank basis. & $C_e$ & Private  coefficients. \\
\hline
\end{tabular}
\end{table}
The common autoregressive inference consists of prompt prefill followed by iterative decoding. Prefill encodes the input prompt, establishing the context for decoding, whereas decoding generates the output continuation one token at a time. We focus on decoding instead of prefill in this paper. This is because the prefill phase processes many tokens in parallel, leading to a compute-bound (instead of memory-bound) workload, while the decoding phase typically generates tokens one by one, and hence is memory-bound. 
The decoding phase can take more than half of end-to-end inference latency, when the output sequence length is long enough\cite{pope2023inference, patel2024splitwise}.


We do not study the attention mechanism. This is because in modern MoE models, the attention accounts for much less weights and much shorter inference latency, compared with the expert computation. For example, the attention accounts for only 3.03\% of overall weights and 10.4\% of end-to-end decoding latency in Qwen3-30B-A3B on AWS Trainium3. 

\subsection{Scratchpad-Based Tensor Accelerator}
\label{sec:bg-hw}

We study STA, commonly characterized with two properties. (1) There is an explicitly managed on-chip SRAM (i.e., scratchpad) for staging and reusing tensors. This SRAM is not managed by hardware. At the software level, The compute kernel or compiler controls tensor allocation and lifetime in the scratchpad. (2) There are multiple heterogeneous compute engines, usually including tensor engine  along with programmable vector, scalar, or SIMD units (or engines). Those engines can work in parallel. 


\begin{figure}[t]
  \centering
  \includegraphics[width=\columnwidth]{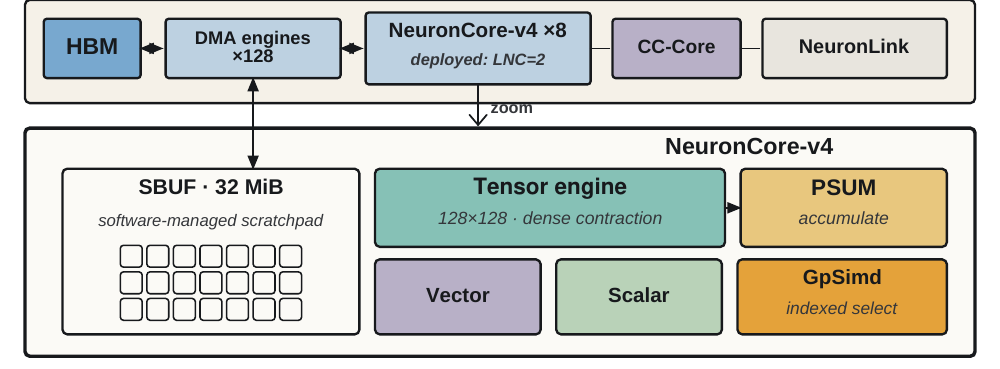}
  \caption{A NeuronCore-v4 in Trainium3.}
  \Description{The diagram connects HBM, DMA engines, NeuronCores, and NeuronLink. A magnified NeuronCore contains an SBUF scratchpad, a tensor engine with PSUM accumulation, and vector, scalar, and programmable SIMD engines.}
  \label{fig:npu-architecture}
\end{figure}



\textbf{AWS Trainium3.} We use AWS Trainium3 as an example of STA. A Trainium3 device contains eight NeuronCore-v4 cores and 144\,GiB of HBM with 4.7\,TB/s bandwidth~\cite{aws2026trainium3arch}. Each NeuronCore-v4 has an SRAM-based 32\,MiB SBUF (i.e., the SPM in STA), an SRAM-based 2\,MiB  accumulator (called PSUM), and 16 DMA engines that move tensors between HBM and SBUF~\cite{aws2026nkidmaoverview}.

\begin{figure*}[t]
  \centering
  \includegraphics[width=\textwidth]{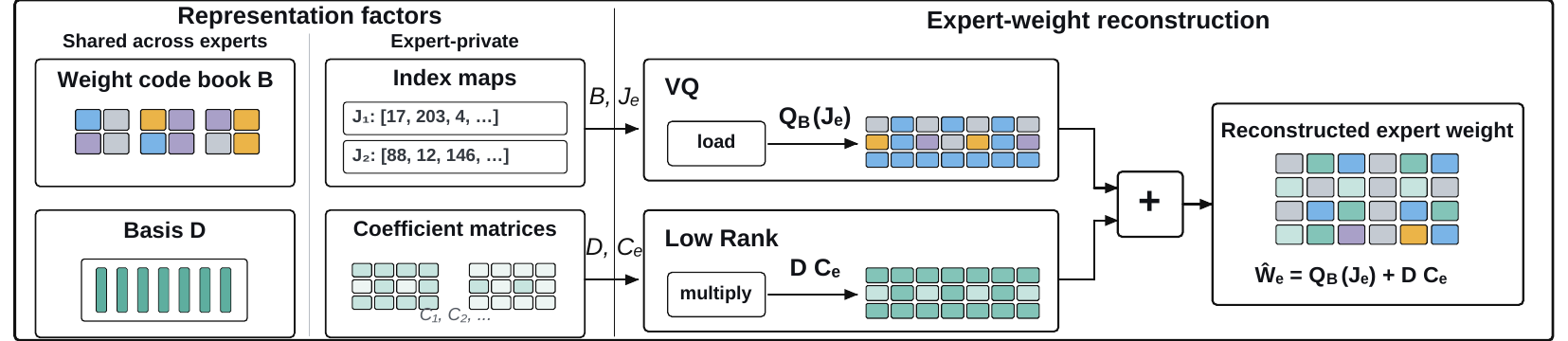}
  \caption{Hybrid expert-weight representation. Shared $\mathcal B$ and $D$ combine with expert-private $J_e$ and $C_e$ to form VQ and LR whose sum represents $\widehat W_e$.}
  \Description{The expert weight is represented as the sum of a vector-quantized base and a low-rank correction. The code book and low-rank basis are shared across experts; index maps and correction coefficients are private to each expert.}
  \label{fig:method-intuition}
\end{figure*}


Each NeuronCore-v4 has four compute engines: a tensor engine, a vector engine, a scalar engine, and a GpSimd engine. The tensor engine is a $128\times128$ grid (or systolic array) of multiply-accumulate units for dense matrix multiplication (matmul), and each matmul holds one matrix stationary in the grid and streams the other through it; the matmul results can be accumulated in PSUM. The vector and scalar engines perform fixed element-wise operations such as expert activation. The GpSimd engine runs programmable SIMD operations that cannot be efficiently mapped to other specialized engines. We program the above hardware resources through Neuron Kernel Interface (NKI) which exposes engine-specific instructions, explicit HBM--SBUF DMA, and SBUF and PSUM.

\subsection{Expert Weight Representations}
\label{sec:bg-compression}

The existing efforts use the LR and VQ \textit{to save memory}. We review them as follows.

\noindent\textbf{Low rank (LR) method} factorizes a given matrix to two lower-rank matrices. In the context of MoE, the LR can approximate the expert-weight matrix of all experts in a layer (i.e., $W$) as a product of two low-rank matrices: $W \approx DC$, where $D \in \mathbb{R}^{H\times r}$ is the basis tensor (rank-$r$) and shared across the experts in the same layer and $C$ is a collection of expert-private coefficients~\cite{molae2025,chen2025mobe}. As a result, for each expert $e$, its weight matrix $W_e \approx DC_e$ where $C_e \in \mathbb{R}^{r\times I}$ is an expert-private coefficient matrix. $C_e$, while preserving expert-private information,  cannot recover information discarded by the shared projection $D^\top$.

\textbf{Vector quantization (VQ)} is built upon a set of representative vectors, each of which is called a code word. The vector set is called the code book containing $K$ code words. Given a vector, VQ approximates it by one of the code words in the code book. Hence, instead of storing the vector itself, VQ stores only an integer index that points to a specific code word in the code book. As a result, the vector is recovered by a single table lookup~\cite{linde1980vq}.

We apply a blockwise form of VQ to the expert weight matrix of all experts~\cite{jegou2011pq}: each column of the expert-weight matrix is partitioned into blocks, each of which is length-$d$, and each weight block is replaced by a code word. Given a weight block, the position (or index) of its code word in the code book (called the \textit{weight code book} in the remaining of the paper) is stored in an index map. Experts in the same layer share the weight code book but keep separate index maps, allowing each expert to use a different combination of code words. 
A larger $d$ reduces the numbers of weight blocks and indices, but generally increases approximation error because each code word represents a larger block. 

\section{Hybrid Expert-Weight Representation}
\label{sec:method}


\name combines LR and VQ to reduce critical-path expert-weight traffic without degrading model quality or creating large runtime overhead. 

\subsection{Hybrid Expert-Weight Representation}
\label{sec:m-form}




\name represents each weight matrix of gate and up projection as the sum of LR and VQ. In the existing efforts, both LR and VQ were individually proposed to shrink model memory (Section~\ref{sec:bg-compression}); \name combines them for a different purpose --- reducing expert-weight movement on the critical path of decoding on STA and increasing utilization of the heterogeneous compute engines. In \name, given $W_e\in\mathbb R^{H\times I}$ and input activation $h$, the expert computation performs the following.
\begin{equation}
  \label{eq:forward}
  o_e=W_e^{\top}h
  \approx\widehat W_e^{\top}h
  =\underbrace{Q_{\mathcal B}(J_e)^{\top}h}_{\text{VQ}}
  +\underbrace{C_e^{\top}(D^{\top}h)}_{\text{LR}}.
\end{equation}
where $J_e$ is expert $e$'s private index map into the weight code book, and $Q_{\mathcal B}(J_e)$ is the resulting $H\times I$ VQ matrix obtained from these weight code book lookups. 
The weight code book $\mathcal B$ and LR basis tensor $D$ are shared across experts within each layer and projection: 
within a layer, the gate and up-projection weights have separate weight code books $\mathcal B$, and have the LR's basis tensor $D$ shared across experts. Meanwhile, each expert holds private $J_e$ and $C_e$ for each of the gate and up projections. Section~\ref{sec:m-motivation} discusses why the LR and VQ complement each other. During decoding, the layer-shared $\mathcal B$ and $D$ are loaded once per token, and each activated expert adds only its $J_e$ and $C_e$ to the expert-weight payload. 

We do not apply LR and VQ to the down projection and keep the down-projection weight matrix in its original BF16, because of two reasons. First, all activated experts apply their gate and up projection weights to the same $h$, allowing the work involving the shared $\mathcal B$ and $D$ to be reused; the down projection weights receive 
expert-specific intermediate activation $p_e$ from gate and up projection.  
Applying \name{}'s VQ + LR representation to the down-projection weights would therefore multiply $\mathcal B$ and $D$ by a different $p_e$ for every activated expert.
Second, over 80\% of the DMA time for down-projection weights overlaps gate/up computation, leaving little on the critical path.


\subsection{Representation Complementarity} 
\label{sec:m-motivation}

\begin{figure}[t]
  \centering
  \includegraphics[width=\columnwidth]{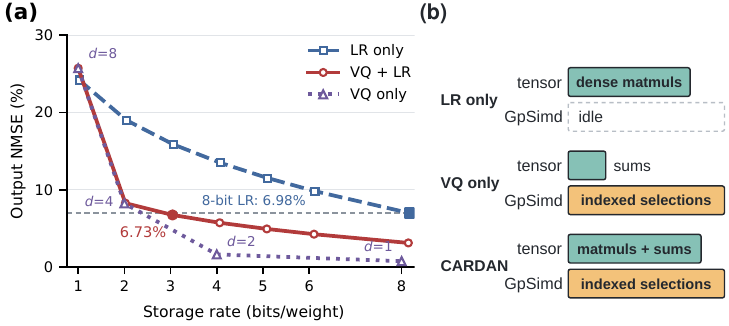}
  \caption{The accuracy of hybrid weight representation for Qwen3-30B-A3B, and work placement. (a) Gate/up-projection output NMSE versus storage rate, averaged over Layers 0, 24, and 47. (b)  engines that executes each representation.}
  \Description{The left panel compares gate and up projection output error against storage rate for Qwen3 representations. The right panel maps dense low-rank computation and indexed vector-quantized selection to different accelerator engines.}
  \label{fig:spectral-motivation}
\end{figure}

The LR and VQ complement each other in \name.
We use Qwen3-30B-A3B (48 layers in total) as an example to depict why \name needs both.
We quantify output quality using the metric, normalized mean squared error (NMSE) at gate and up projections of experts at Layers 0, 24, and 47.
In our study, NMSE measures the difference between a weight representation (LR only, VQ only, or LR+VQ) and the BF16 output (the original output), defined as follows.  
\begin{equation}
    \label{eq:nmse}
    \mathrm{NMSE}=\frac{\sum\|\widehat W_e^{\top}h-W_e^{\top}h\|_2^{2}}{\sum\|W_e^{\top}h\|_2^{2}},
\end{equation}
where $\widehat W_e$ is the weight reconstructed by LR only, VQ only, or VQ+LR, and the summation terms run over evaluation tokens $h$, experts, and both gate and up projection within a layer, and report the mean of the three per-layer NMSE values for Layers 0, 24, and 47. 
A lower NMSE indicates that the output is closer to the original BF16 output, and vice versa. Figure~\ref{fig:spectral-motivation}(a) plots NMSE against storage rate (i.e., the number of bits used per weight). 



Figure~\ref{fig:spectral-motivation}(a) compares the three representations. At approximately 4 bits/weight, NMSE is 13.46\% for LR only, 5.72\% for VQ+LR, and 1.62\% for VQ only ($d=2$). LR only at 8 bits/weight (6.98\%) is still worse than VQ+LR at 3 bits/weight (6.73\%), so VQ+LR needs less than half the storage for the same accuracy. VQ only is more accurate, but only by shrinking $d$, and every halving of $d$ doubles the indexed selections per weight, making indexed selection a major overhead on critical path. 



\noindent\textbf{Workload placement on STA.} The LR and VQ are also complementary in terms of the placement on heterogeneous compute engines, figure~\ref{fig:spectral-motivation}(b) shows this complementarity.  
The LR computes $z=D^\top h$ once per input and $C_e^\top z$ for each activated expert using dense matrix--vector multiplications on the tensor engine. The VQ uses each expert's indices $J_e$ to select code words from the projection code book in SBUF (Section~\ref{sec:i-shared}):  one indexed selection per weight block. The indexed selections for weights are independent from each other and use less compute, hence performed on GpSimd. In addition, the LR computation and VQ lookup happen in parallel  because neither requires the other's output.

After the gate/up-projection outputs are formed, the scalar engine applies model-specific expert activation $\phi$ to the gate-output vector. The vector engine then multiplies corresponding elements of the activated gate-output vector and the up-projection-output vector, producing the intermediate activation $p_e$ for the down projection. 
\subsection{Initialization of Hybrid Representation}

\begin{figure}[t]
  \centering
  \includegraphics[width=\columnwidth]{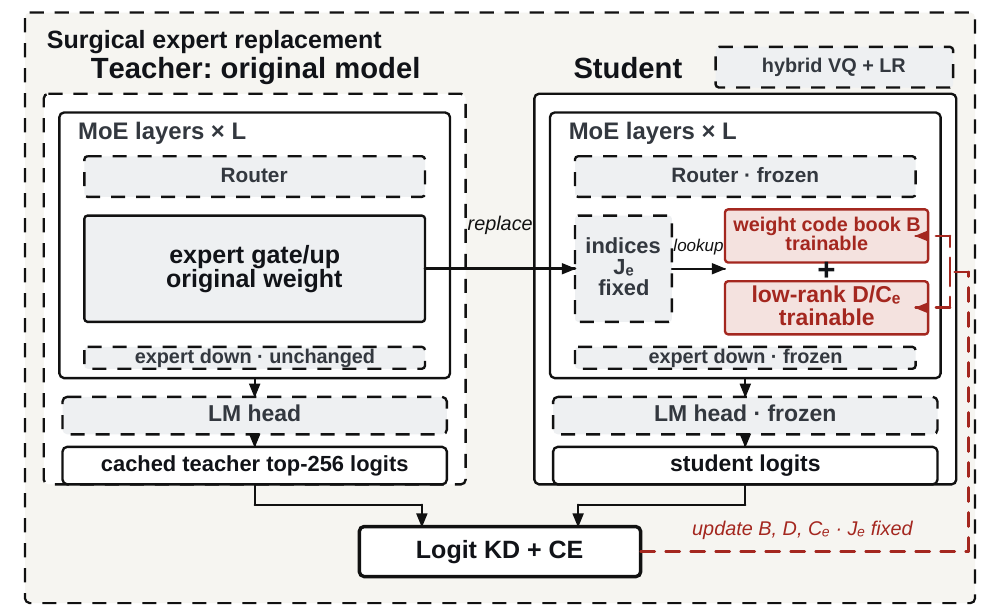}
  \caption{Training scope. Only the gate/up weight code book, LR basis tensor $D$ and coefficient matrix $C_e$ are trained.
  }
  \Description{Teacher and student models share the model structure. The student replaces expert gate and up weights with a hybrid representation. Distillation and cross-entropy update the code book, basis, and coefficients while indices and the remaining model parameters stay fixed.}
  \label{fig:factor-preparation}
\end{figure}

Building the hybrid representation must initialize the weight code book $\mathcal B$, the index maps $J_e$, the LR basis tensor $D$, and the coefficient matrices $C_e$, so that the output of expert computation and the final output of the model match those of the original output in BF16. 
\name first applies a training-free, layer-wise initialization to obtain a working representation without any gradient training. A short end-to-end distillation then restores the model quality to the BF16 level. (Figure~\ref{fig:factor-preparation}).

\textbf{Training-free initialization.} We initialize the VQ using the standard product quantization~\cite{jegou2011pq}, applied separately to the gate and up weights of each layer. This initialization produces the shared weight code book per layer $\mathcal B$ and the expert-private index maps $J_e$~\cite{jegou2011pq}. Then, we fit the LR in closed form on representative decode activations, so the fit prioritizes the errors that affect expert outputs~\cite{svdllm2025}. The fit obtains one $D$ shared across experts in a  layer and a separate $C_e$ for each expert, following standard LR correction methods~\cite{yao2023zeroquantv2,saha2024caldera,zhang2025qera}. Using the closed-form fit is beneficial, because it provides a usable initialization without gradient-based training or learning-rate tuning. Appendix~\ref{app:details} formalizes the objective for a layer and derives its solution.


\textbf{End-to-end distillation.} The layer-wise initialization minimizes each layer's own output error, but it cannot see how errors are accumulated and propagated across the layers. To solve this problem,  we use knowledge distillation (KD) that jointly refines all layers~\cite{egiazarian2024aqlm,tseng2024quip,liu2024vptq}. 
In particular, the teacher is the original BF16 model with all weights frozen, and the student is the same model with the hybrid representation in place of the expert gate and up weights.
\begin{figure*}[t]
  \centering
  \includegraphics[width=\textwidth]{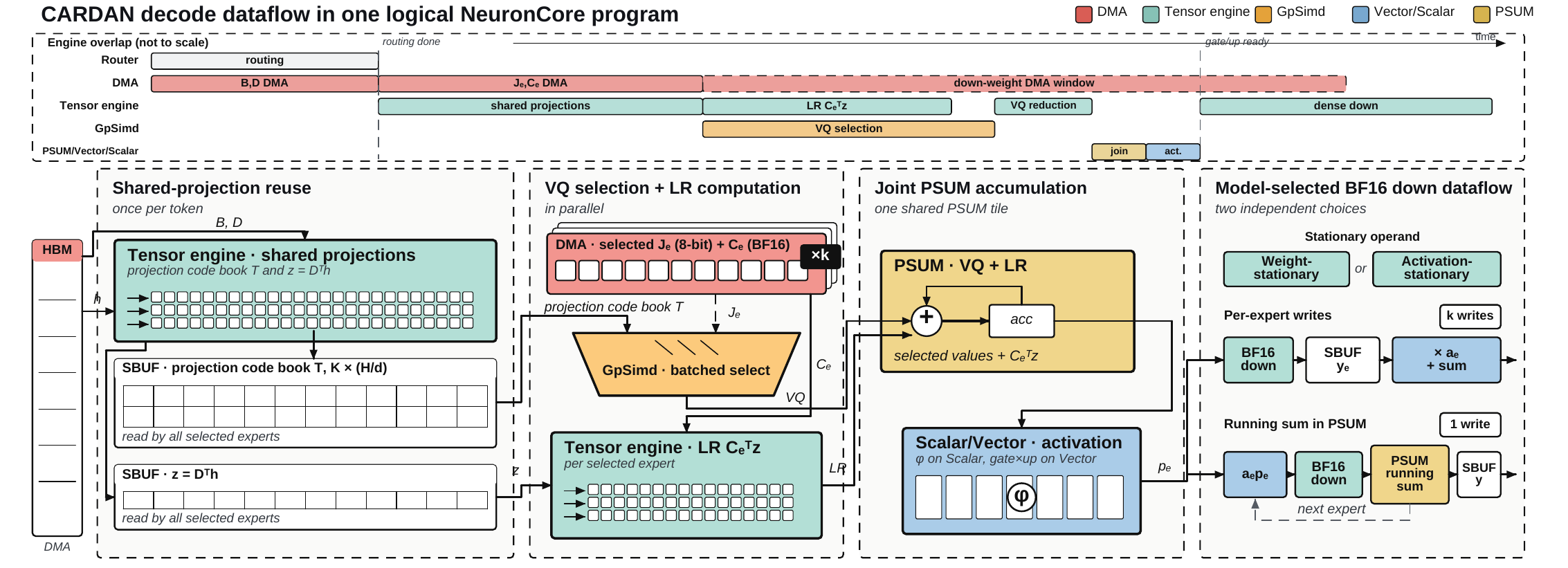}
  \caption{\name's decoding dataflow. Top: $B$,$D$ data transfer overlaps routing, expert-private $J_e$,$C_e$ data transfer overlaps shared projections, down projection weight data transfer overlaps the GpSimd and tensor engine gate/up computation.
  Bottom: the mechanisms enabling these overlaps.}
  \Description{A schematic timeline places routing, shared-factor transfers, shared projections, expert-private transfers, indexed selection, low-rank computation, and down projection on concurrent engines. Lower panels illustrate shared-projection reuse, joint accumulation, and alternative down-projection schedules.}
  \label{fig:implementation-dataflow}
  \vspace{-3pt}
\end{figure*}
For each token, $\Omega$ denotes the set of vocabulary indices of the teacher's 256 largest logits. We take the teacher and student logits at the indices in $\Omega$, divide both by a scaling factor $\tau$, and apply softmax, yielding $p_{\mathrm{tea}}^{\Omega}$ and $p_{\mathrm{stu}}^{\Omega}$.
With $\mathcal L_{\mathrm{CE}}$ denoting next-token cross-entropy, we minimize the following.

\begin{equation}
\label{eq:kd-loss}
  \mathcal L
  = \tau^2 D_{\mathrm{KL}}
    \!\left(p_{\mathrm{tea}}^{\Omega}\,\|\,p_{\mathrm{stu}}^{\Omega}\right)
  \;+\; \lambda\,\mathcal L_{\mathrm{CE}}.
\end{equation}

\noindent where $\lambda$ is a model-quality alignment hyperparameter. 
During KD, only $\mathcal B$, $D$, and $C_e$ in the student are updated; $J_e$ and all other model parameters are frozen (Figure~\ref{fig:factor-preparation}). 

\section{Multi-Engine Decoding Dataflow}
\label{sec:impl}

The weight representation in Section~\ref{sec:method} reduces memory traffic and introduces work across multiple compute engines. \name organizes the decoding work into a multi-engine dataflow which breaks the routing--transfer--compute dependency chain and overlaps data movement with computation. Figure~\ref{fig:implementation-dataflow} summarizes the resulting dataflow. 
\subsection{Multi-Engine Overlap Schedule}
\label{sec:i-overlap}

Section~\ref{sec:m-motivation} places the LR tensor multiplications on the tensor engine and the VQ selections on GpSimd. This subsection describes \emph{when} that work and the associated DMA transfers are issued, so that the DMA transfer is hidden along the routing--transfer--compute chain (Figure~\ref{fig:implementation-dataflow}, top).
\name separates routing-independent shared work (i.e., loading $\mathcal B$ and $D$ and multiplying them by $h$) from routing-dependent expert-private work (i.e., loading and applying $J_e$, $C_e$, and the down projection). This separation exposes three windows to overlap the data transfer and compute.



\begin{itemize}
    \item The shared code book $\mathcal B$ and LR basis tensor $D$ do not depend on which experts are activated, so the DMA transfers of $\mathcal B$ and $D$ can be issued while the router is computing the expert scores and selecting top-$k$ experts.

    \item After the routing, the tensor engine performs two operations: (1) the rank-$r$ projection $z=D^\top h$ and (2) multiplying each length-$d$ block of $h$ with $K$ code words of $\mathcal B$. We call the two operations \textit{the shared projections}. Meanwhile, DMA loads the activated experts' $J_e$ and $C_e$.

    \item GpSimd performs the $J_e$-indexed selections while the tensor engine computes $C_e^\top z$. The addresses of the selected down-projection weights are known after the routing, and do not depend on the gate/up-projection outputs, so the DMA transfer of those weights can be issued during the computation on the GpSimd and tensor engine.
\end{itemize}





\subsection{Removing Redundancy in Compute and Memory}
\label{sec:remove_redunt}
\phantomsection\label{sec:i-shared}
\phantomsection\label{sec:i-gather}
\phantomsection\label{sec:i-fuse}
The multi-engine dataflow has redundancy in compute and memory: (1) A straightforward execution of the hybrid representation would reload $B$ and $D$ and recompute the shared projections for each expert. (2) In VQ, GpSimd uses each expert's index map to access the code book.
GpSimd repeatedly pays the setup cost of issuing selection instructions, leading to redundancy in GpSimd compute. (3) The parallelization of LR and VQ leads to independent stores of the LR/VQ results in SBUF, leading to memory redundancy during the summation of the LR and VQ results. We use the following techniques to remove the redundancy.


\textbf{Reuse of shared projections.} \name performs the shared projections only once \textit{before} the expert computation, rather than once per expert during the expert computation. As a result, each expert can reuse the shared projections, removing the repeated loading of $D$ and $B$, and matmul for the projection.  In particular, before the expert computation, the tensor engine performs $z = D^T h$ for LR, to be shared between experts; also, the tensor engine multiplies every length-$d$ block $h_b$ of $h$ with $K$ code words in the weight code book $\mathcal B$, producing a \textit{projection code book} $T[b,k]=h_b^{\top}\mathcal B_k$. During the expert computation, instead of using the weight code book, an expert selects code words from the projection code book $T$ to select code words.




Both $z$ and $T$ stay in SBUF from the stage 1 until all activated experts have used them (see Figure~\ref{fig:implementation-dataflow}). 
The book $T$ is small. It holds $KH/d$ values (in BF16) per gate- or up-projection. 
For example, with $K{=}256$ and $H{=}2048$, the two projection-code-books for the gate- and up-projection together occupy only $0.5$\,MiB at $d{=}4$ or $0.25$\,MiB at $d{=}8$. 


\textbf{Cross-expert gather batching.} The VQ uses each expert's index map $J_e$ to pick code words from $T$. A straightforward implementation would issue one indexed gather per activated expert. Since each gather pays a constant setup cost on GpSimd, the setup costs of all activated experts add up on the critical path. 

\name concatenates $J_e$ of the activated experts 
and passes the concatenation as a batch to the NKI ISA primitive \texttt{nc\_n\_gather}. Within each of the gate and up projections, since the activated experts select from the same $T$ concurrently using their own index maps, batching preserves every lookup while paying the constant gather setup once per batch, rather than once per expert. Furthermore, because $J_e$ is loaded separately for every activated expert, \name stores the indices in $J_e$ as \texttt{uint8}, which limits expert-private HBM traffic and addresses exactly the $K{=}256$ code words. This solution is depicted in the stage~2 in Figure~\ref{fig:implementation-dataflow},

\begin{figure}[!t]
  \centering
  \includegraphics[width=\columnwidth]{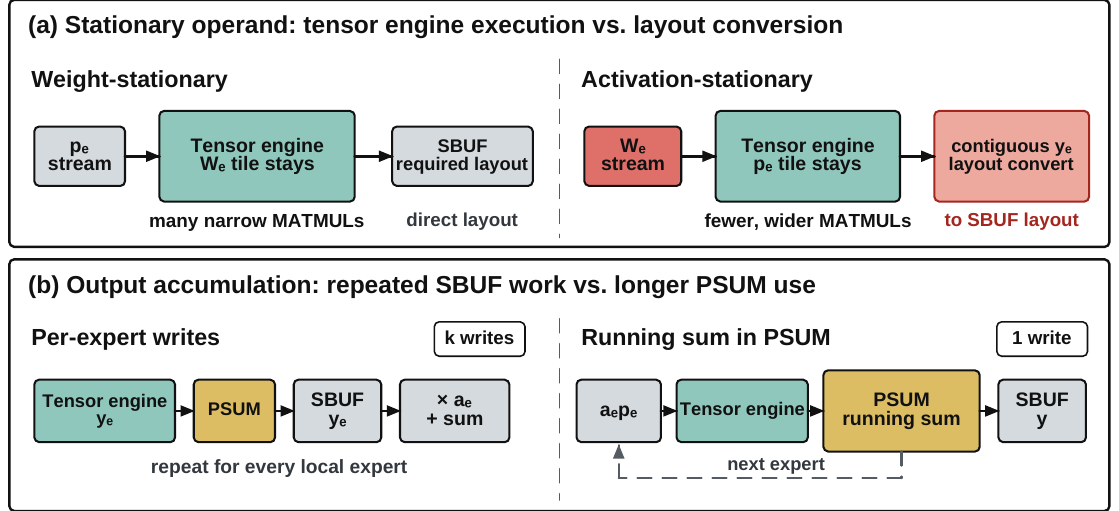}
  \caption{BF16 down projection scheduling choices: (a) weight- versus activation-stationary execution; (b) per-expert SBUF writes versus cross-expert PSUM accumulation.}
  \Description{Two pairs of diagrams compare down-projection schedules. Weight-stationary and activation-stationary execution differ in matrix multiplication and layout conversion. Per-expert SBUF writes and a running sum in PSUM trade repeated output writes against longer accumulator use.}
  \label{fig:down-schedules}
  \vspace{-10pt}
\end{figure}
\textbf{Joint accumulation of LR and VQ.} The LR and VQ reach the tensor engines in different forms. The VQ provides gathered code words that must be reduced, whereas the LR computes the dense product $C_e^\top z$.
Their sum is the gate or up projection output $o_e$ in Eq.~\eqref{eq:forward}, and the expert activation consumes only this summation. Materializing the LR and VQ independently would create separate spaces in SBUF, and require an additional read-add-write before activation. As shown in the stage~3 in Figure~\ref{fig:implementation-dataflow}, \name maps the addition of the LR and VQ onto Trainium3's PSUM for accumulation. Specifically, the tensor engine sums the code words selected from $T$ in PSUM, and then accumulates the LR matmul $C_e^{\top}z$ into the same PSUM locations as the code-word summation. 
Neither LR nor VQ needs to write a standalone output vector to SBUF, reducing memory redundancy in SBUF.

\subsection{Schedule of Down Projection}
\label{sec:i-down}

The down projection computes $y_e=W_{\mathrm{down},e}^{\top}p_e$ \hspace{2pt} per activated expert $e$ and $\sum_e a_e y_e$ (named the \textit{accumulated summation}) across the activated experts. After the optimizations in Section~\ref{sec:remove_redunt} on the gate- and up-projections, the down-projection, dominated by dense matmul, dominates the remaining time on the tensor engine.

There are two independent dimensions to control the schedule of the down projection. (1) For each activated expert's down-projection matmul ($y_e=W_{\mathrm{down},e}^{T} p_e$),  
which input ($W_{\mathrm{down},e}$ or $p_e$) should be held stationary in the tensor engine's systolic array while the other streams through it for better performance? 
This dimension impacts how many matmuls are issued and whether the data-layout conversion is needed after matmul.  (2) Where is the accumulated summation operated? On SBUF or PSUM? If on SBUF, as soon as  $y_e$ for an expert is produced on PSUM, $y_e$ is copied to SBUF and accumulated there, which frees PSUM immediately but costs an extra SBUF write per expert. 
If on PSUM, the accumulated summation accumulates $y_e$ on PSUM and is then written to SBUF after \textit{all} activated experts are done. Compared with on SBUF, on PSUM, the accumulated summation takes a longer time in PSUM, making PSUM less available to the gate/up projections for better performance.

The two dimensions give four candidate schedules. We decide which schedule performs best using the offline profiling. 
The offline profiling is feasible. In particular, once an MoE model is selected, the shapes and layout of the down projection, which is needed for deciding the schedule, is determined and do not change across tokens and experts. Hence, the offline profiling results can be used to determine the schedule for online inference. We describe our scheduling decision as follows. 

\textbf{Deciding stationary input.} See Figure~\ref{fig:down-schedules} (a). The execution of weight-stationary generates many skinny matmuls, but directly produces the required SBUF layout. Skinny matmuls come from the multiplication of $p_e$ streaming with the weight tiles resident in the systolic array; the skinny matmul results in the systolic array can be directly used as the down projection without the need of data-layout conversion. 
The execution of the activation stationary uses fewer, wider matmul, but must convert its output to be aligned with the data-layout of the down projection. 
\name uses the activation stationary when the tensor engine time the activation stationary saves on the critical path exceeds the layout-conversion time (e.g., the models OLMoE and Nemotron); otherwise, the weight stationary is employed (e.g., the models Qwen3-30B-A3B, DeepSeek-V2-Lite, and Gemma-4-26B-A4B). 

\textbf{Deciding the location of the accumulated summation.} See Figure~\ref{fig:down-schedules}(b).  
Keeping the accumulated summation on  PSUM pays off when $y_e$ copying and accumulating on SBUF this method removes is on the critical path (e.g., see the models DeepSeek-V2-Lite, Nemotron, and Gemma~4). When $y_e$ copying and accumulating on SBUF are already hidden behind other operations (e.g., see the models Qwen3-30B-A3B and OLMoE), \name writes $y_e$ to SBUF. 

\section{Evaluation}
\label{sec:eval}
\name is implemented as an open-source project with 9.7K lines of code. We evaluate \name with multiple MoE models summarized in Table~\ref{tab:eval-setup}. 

\subsection{Experimental Setup}
\label{sec:e-setup}



All performance experiments run on AWS Trainium3 with four-way tensor parallelism (TP${=}4$), using eight physical NeuronCore-v4s in total. We use logical NeuronCore configuration 2 (LNC${=}2$), which groups two physical NeuronCores into one logical NeuronCore. All models use Neuron SDK 2.31 (neuronx-cc 2.26, NKI 0.5).


We measure batch-one autoregressive decoding unless stated otherwise (Section~\ref{sec:e-batch}), excluding compilation and model loading. 
We report time per output token (TPOT). For each configuration, we generate 19, 64, and 110 tokens from the same prompt, take the median host time of 24 runs after two warmups at each length, and fit latency as a linear function of generated tokens. TPOT is the slope of this line. Confidence intervals (CIs) are the 2.5th and 97.5th percentiles of TPOT over 10,000 bootstrap resamples of these runs.
For device profiles, \emph{device-graph span} denotes the elapsed time from the first to the last operation in a decode-step graph per token.



We compare \name against two baselines.

\begin{itemize}
    \item \textbf{PyTorch/XLA.} This compiler-lowered full-model path preserves the original dense BF16 expert computation and serves as our first baseline without any  custom MoE kernel.
    
    \item \textbf{Dense megakernel.} This optimized MoE implementation based on BF16  is from AWS and works as our primary performance baseline. For Qwen3-30B-A3B and Gemma-4-26B-A4B we use the production megakernel from the public AWS NeuronX Distributed Inference (NxDI) release. Because NxDI fuses MoE decode only for Qwen3-30B-A3B and Gemma-4-26B-A4B, we port the same kernel design to the other three models to obtain a consistent optimized baseline, adapting expert dimensions, activation, and surrounding layers while keeping BF16-based expert computation. 
\end{itemize}




\begin{table}[t]
\centering
\caption{MoE models for evaluation.}
\label{tab:eval-setup}
\footnotesize
\renewcommand{\arraystretch}{0.94}
\setlength{\tabcolsep}{2.4pt}
\begin{tabular}{@{}lccccc@{}}
\hline
model & $L/L_{\mathrm{MoE}}$ & $H/I$ & $E/k$ & proj. & $(d,K,r)$ \\
\hline
Qwen3-30B-A3B  & 48/48 & 2048/768  & 128/8 & gate+up & $(8,256,384)$ \\
DeepSeek-V2-Lite  & 27/26 & 2048/1408 & 64/6  & gate+up & $(8,256,384)$ \\
Nemotron-3-Nano-30B   & 52/23 & 2688/1856 & 128/6 & up      & $(8,256,512)$ \\
OLMoE-1B-7B  & 16/16 & 2048/1024 & 64/8  & gate+up & $(4,256,256)$ \\
Gemma-4-26B-A4B & 30/30 & 2816/704  & 128/8 & gate+up & $(8,256,384)$ \\
\hline
\end{tabular}
\end{table}



We measure quality by perplexity (PPL) on WikiText-103, evaluated on 32 windows of 512 tokens that are disjoint from the 10,000 256-token windows used for factor initialization and distillation. We report each model against its own BF16 teacher measured identically.




\subsection{End-to-End Performance} 
\label{sec:e-gen}

\begin{figure}[t]
\centering
\includegraphics[width=\columnwidth]{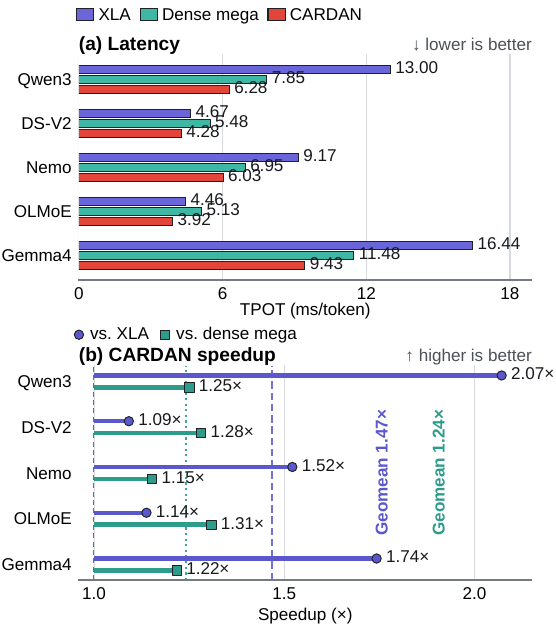}
\caption{Full-model batch-one decode for deployed configurations:
(a) host-observed TPOT; (b) \name{} speedup (lines: geometric means).
Qwen3-30B-A3B and Gemma-4-26B-A4B use the production AWS/NxDI dense kernel; The other baselines port its design.}
\Description{Two panels compare latency bars and CARDAN speedup markers against XLA and dense megakernels across five MoE models. Reference lines mark geometric-mean speedups.}
  \label{fig:e2e}
\end{figure}

Figure~\ref{fig:e2e} reports end-to-end results.  Across the five models, \name{} is $1.09$--$2.07\times$ faster than PyTorch/XLA and $1.15$--$1.31\times$ faster than the dense megakernel.  The corresponding geometric-mean speedups are $1.47\times$ and $1.24\times$. 
DeepSeek-V2-Lite and Nemotron-3-Nano-30B-A3B additionally fuse dense projections that read the same hidden state into one wider matmul. Section~\ref{sec:e-ablation} separates the gain of this fusion.
These gains come from three sources. The hybrid representation reduces the expert-weight data movement that dominates decode. Shared projections are computed once per token instead of once per expert. The overlap schedule hides the remaining transfers and the added indexed selections behind concurrent engine work, keeping their overhead off the critical path.



\subsection{Workload Placement and Critical-Path Analysis}
\label{sec:e-sys}

\begin{figure}[t]
\centering
\includegraphics[width=\columnwidth]{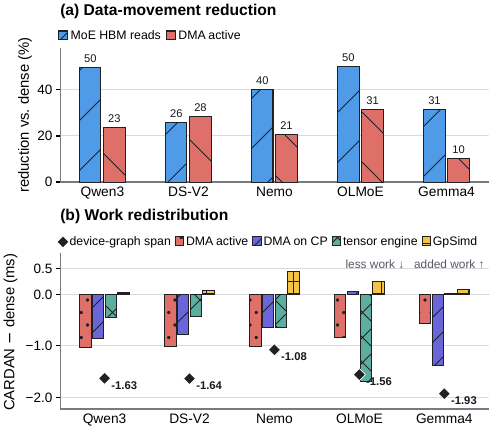}
\caption{Full-model changes from the dense megakernel: (a) reductions in MoE
HBM reads and DMA activity/service; (b) engine-union deltas, with diamonds
marking device-graph-span deltas. Intervals overlap.}
\Description{The upper panel shows reductions in MoE memory reads and DMA activity. The lower panel shows changes in DMA, tensor-engine, and programmable SIMD active intervals, with diamonds marking the change in device-graph duration. Engine intervals can overlap.}
  \label{fig:engine-breakdown}
\end{figure}

We compare \name{} with the dense megakernel, the baseline which is already a tuned MoE kernel, so the difference isolates the representation and its dataflow. For each model we profile the deployed configuration of Section~\ref{sec:e-setup} over all layers, running the decode-step graph five times and selecting the trace closest to the median device-graph span. Figure~\ref{fig:engine-breakdown} reports one rank of the TP${=}4$ execution. DMA-active time is the total duration during which at least one physical DMA packet is active.

\noindent\textbf{DMA reduction.}
Figure~\ref{fig:engine-breakdown}(a) shows that \name{} reduces MoE-projection HBM reads by 26--50\% and DMA-active time by 10--31\% across the five models. Total HBM reads fall by 10--32\%.

\begin{table}[t]
\centering
\caption{Gemma-4-26B-A4B transition from \name{} to equal-rate VQ-only. HBM is MB/rank; other values are ms. D/P/G denote DMA service, tensor engine, and non-DMA GpSimd. The final D/G column reports work not hidden by another engine.}
\label{tab:qwen-engine-boundary}
\footnotesize
\renewcommand{\arraystretch}{0.92}
\setlength{\tabcolsep}{1.5pt}
\begin{tabular*}{\columnwidth}{@{\extracolsep{\fill}}lrrrrrr@{}}
\hline
representation & HBM & \shortstack{device\\span} &
\multicolumn{3}{c}{total union} & \shortstack{non-overlap\\D/G} \\
& & & D & P & G & \\
\hline
\name{} $(8,512)$ & 1834 & 8.290 & 6.182 & 2.284 & 0.687 & 3.287/0.123 \\
VQ only $(2,0)$ & 1763 & 9.867 & 6.521 & 2.559 & 2.053 & 3.510/0.852 \\
\hline
\end{tabular*}
\vspace{-5pt}
\end{table}
\noindent\textbf{Workload redistribution.}
Figure~\ref{fig:engine-breakdown}(b) compares \name{} with the dense megakernel. \name{} adds $0.04$--$0.45$\,ms of GpSimd work but removes $0.84$--$1.04$\,ms of DMA-active time and $0.42$--$1.69$\,ms of tensor-engine time (per-engine occupancies, not critical-path time, as the engines overlap).  The added GpSimd selections run concurrently with the remaining DMA and tensor-engine work, the device-graph span falls by $1.08$--$1.93$\,ms across the five models.

\noindent\textbf{Traffic versus runtime overhead.}
Each VQ block needs one indexed selection on GpSimd, so reducing $d$ increases selection count and can offset the DMA savings.
For Gemma-4-26B-A4B, we compare \name{} at $(d,r)=(8,512)$ with VQ only at $(d,r)=(2,0)$, both at $4.00$ bits/weight.
The $r=512$ setting is used only for this equal-storage ablation.
Setting $r=0$ removes the LR. To match \name{}'s storage rate, VQ only reduces $d$ from 8 to 2, which increases the number of indexed selections by $4\times$. Relative to 
equal-storage \name{}, VQ only reads 3.9\% fewer HBM bytes but increases GpSimd time by 199\%. 
The GpSimd time exposed on the critical path rises from $0.123$\,ms in \name{} to $0.852$\,ms in VQ only, more than the exposed time of any other compute engine. 
Device-graph span rises by 19.0\%, and TPOT rises from 10.571 to 12.162\,ms/token, above the 11.48\,ms/token dense baseline. Thus, at a matched storage rate, fewer transferred bytes do not compensate for the added indexed-selection work.



\subsection{Kernel Design Ablation}
\label{sec:e-ablation}

Table~\ref{tab:kernel-ablation} isolates the benefit of \name{} from fusion of surrounding projections. Part~A adds reuse of shared projections and then batches gathers across experts. Part~B compares dense and \name{} execution for DeepSeek-V2-Lite and Nemotron Nano-30B-A3B, with and without this fusion.


The hybrid representation alone reduces Qwen3-30B-A3B TPOT by $12.1\%$, even when the shared projections are recomputed for every expert.  Shared-projection reuse provides a further $4.9\%$ reduction, and cross-expert gather batching provides another $4.3\%$, for a $20.0\%$ cumulative reduction.  
A six-layer profile of the same three mechanism variants at $(d,r)=(4,256)$
shows where each mechanism saves time. Shared-projection reuse reduces
tensor-active time from $632$ to $562\,\mu$s and DMA-active time from
$705$ to $643\,\mu$s; cross-expert gather batching cuts GpSimd selection
instructions from $385$ to $97$ and GpSimd time from $616$ to $502\,\mu$s.


\begin{table}[t]
\centering
\caption{Full-model attribution: (A) \name{} mechanisms on Qwen3-30B-A3B; (B) dense-to-\name{} latency for DeepSeek and Nemotron with and without surrounding-projection fusion.}
\vspace{-5pt}
\label{tab:kernel-ablation}
\footnotesize
\renewcommand{\arraystretch}{0.92}
\setlength{\tabcolsep}{2.5pt}
\begin{tabular*}{\columnwidth}{@{\extracolsep{\fill}}lcc@{}}
\hline
\multicolumn{3}{@{}l}{\textit{A. \name{} mechanisms on Qwen3-30B-A3B ($d{=}8,r{=}384$), 48 layers}} \\
configuration & TPOT (ms) & step reduction \\
dense megakernel & 7.85 & --- \\
VQ $+$ LR, per-expert shared proj. & 6.90 & $12.1\%$ \\
$+$ shared-projection reuse & 6.56 & $4.9\%$ \\
$+$ cross-expert gather batching & \textbf{6.28} & $\mathbf{4.3\%}$ \\
\hline
\multicolumn{3}{@{}l}{\textit{B. \name{} gain with and without fusion}} \\
& \multicolumn{2}{c}{dense $\rightarrow$ \textbf{\name{}} (ms/token)} \\
model & without fusion & with fusion \\
DeepSeek-V2-Lite & $5.48 \rightarrow \mathbf{4.95}$ & $4.82 \rightarrow \mathbf{4.28}$ \\
Nemotron-3 & $6.94 \rightarrow \mathbf{6.25}$ & $6.73 \rightarrow \mathbf{6.03}$ \\
\hline
\end{tabular*}
\end{table}

On Qwen3-30B-A3B, we also tested the down-projection
running sum of Section~\ref{sec:i-down}. It accumulates the activated experts' down outputs in one PSUM bank and writes the final sum to SBUF once. In isolation, this makes the down step $3.22\times$ faster than the per-expert-copy schedule, but increases 48-layer TPOT from $6.278$ to $6.432$\,ms because PSUM remains occupied across the expert loop. 



Surrounding-projection fusion is an optimization independent of the \name{} representation, and it further accelerates \name{}.  Part~B separates the two gains by measuring the dense baseline and \name{} with and without fusion.  Fusion reduces latency on both sides, by $11.9\%$ (dense) and $13.6\%$ (\name{}) on DeepSeek and by $3.1\%$ and $3.6\%$ on Nemotron-3-Nano-30B-A3B.  \name{} stays $1.11$--$1.13\times$ faster than dense with or without fusion, so its speedup does not depend on fusion.

\subsection{Sensitivity and Operating Boundaries}
\label{sec:e-sensitivity}
\label{sec:e-scope}

We fix $K=256$, the largest code book supported by \texttt{uint8} indices, 
and store the code book and low-rank factors in BF16.
Including both expert-private and amortized shared parameters, the storage rate is
\[
b(d,r)=\frac{8}{d}+\frac{16r}{H}+\frac{16r}{EI}
+\frac{16Kd}{EHI}
\]
bits per original weight (Appendix~\ref{app:storage}).
Appendix~\ref{app:work-vector} gives the corresponding per-engine work counts as functions of $d$, $K$, and $r$.
We evaluate $d\in\{4,8\}$ at approximately 3, 4, and 5 bits/weight,
choosing $r$ to match their storage within 0.02 bits/weight.

\begin{figure}[t]
\centering
\includegraphics[width=\columnwidth]{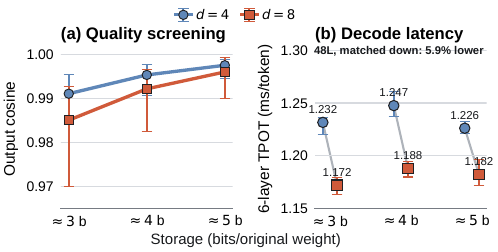}
\caption{Qwen3-30B-A3B sensitivity: (a) gate/up output cosine at Layers 0/24/47;
(b) six-layer TPOT with bootstrap 95\% CIs.}
\Description{Two panels compare Qwen3 configurations with different vector lengths and low-rank dimensions. The first shows projection output cosine similarity at three layers; the second shows six-layer token latency with 95 percent confidence intervals.}
  \label{fig:qwen-sensitivity}
  \vspace{-10pt}
\end{figure}

Figure~\ref{fig:qwen-sensitivity} shows the quality-latency tradeoff. $d{=}8$ lowers TPOT by 3.6--4.8\%, whereas $d{=}4$ retains higher teacher--student output cosine similarity before distillation.
At the 4-bit rate, changing $(d,r)$ from $(4,256)$ to $(8,384)$ lowers GpSimd time from 495 to 440\,$\mu$s despite 4.2\% more HBM reads.  A separate six-layer, equal-rate VQ-only configuration with $(d,r)=(2,0)$ removes all low-rank work and reads 12\% fewer bytes, yet raises GpSimd time from 440 to 603\,$\mu$s and TPOT from about 1.19 to 1.37\,ms/token ($+14.7\%$). 

The full 48-layer VQ-only model with $(d,r)=(2,0)$ takes 8.075\,ms/token. This is 28.6\% slower than \name{} with $(d,r)=(8,384)$ (6.278\,ms/token). It is also slower than the dense BF16 baseline (7.851\,ms/token). With $d=2$, GpSimd selection time exceeds the
overlapping DMA time, shifting the decode bottleneck from DMA to GpSimd.
Table~\ref{tab:qwen-engine-boundary} shows the same shift for
Gemma-4-26B-A4B. Thus, although $d=2$ reduces HBM traffic, its additional
indexed-selection work makes it slower overall, so we use $d=8$. 
We also compare against Trainium's built-in 4-bit-weight, 8-bit-activation (W4A8) quantization, which stores the gate and up weights in the MXFP4 microscaling format while the down projection stays BF16. Under the same full-model TPOT measurement on Qwen3-30B-A3B, W4A8 reaches 11.75\,ms/token, slower than the 7.85 of the BF16 dense megakernel. The slowdown comes from a layout boundary rather than from MXFP4 arithmetic. The MX matmul writes a packed layout that only another MX operation can read, so every activated expert's gate/up output is converted on chip before the BF16 down projection (Appendix~\ref{app:mx-boundary}).

\subsection{Batch Scaling of Selective Decode}
\label{sec:e-batch}

\begin{figure}[t]
\centering
\includegraphics[width=\columnwidth]{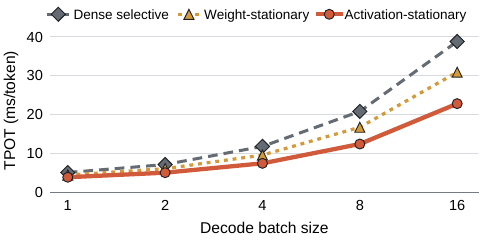}
\caption{Full-model OLMoE-1B-7B batch scaling. Lower is better.}
\Description{A line chart compares token latency as decode batch size increases from one to sixteen for dense selective execution and CARDAN with weight-stationary or activation-stationary down projection.}
  \label{fig:olmoe-batch-scaling}
\end{figure}

We study batch scaling on OLMoE-1B-7B (Table~\ref{tab:eval-setup}), which activates 8 of 64 experts per token. Figure~\ref{fig:olmoe-batch-scaling} measures the full 16-layer model with $(d,K,r)=(4,256,256)$ and the same TPOT method as Figure~\ref{fig:e2e}. Each batch receives fixed, distinct prompts. At every batch size, both dense and \name{} load only the activated experts. 
\name{} remains faster at every batch size, growing from $1.31\times$ at batch size 1 (B1) to $1.67\times$ at B8 and $1.70\times$ at B16. 

Device profiles reveal the sources of \name{}'s performance gains.  
At batch sizes 8 and 16, \name reads 3.33 and 6.13\,GB from HBM,
respectively, compared with 7.02 and 13.56\,GB for the dense baseline. DMA-active time similarly drops from 14.27 to 6.85\,ms at batch size 8 and from 27.46 to 13.14\,ms at batch size 16.
Furthermore, the impact of the down-projection schedule (Section \ref{sec:i-down}) becomes more pronounced as the batch size increases. Because both schedules require nearly identical HBM reads, superior operand placement and tensor-engine scheduling enable the activation-stationary schedule to consistently outperform the weight-stationary variant. This speed advantage grows from 12.5\% at batch size 1 to 26.2\% at 16.

\subsection{Model Quality, Compression, and Kernel Fidelity}
\label{sec:e-quality}
\phantomsection\label{sec:e-comp}
\phantomsection\label{sec:e-kd}

Table~\ref{tab:quality-generalization} reports perplexity and storage reduction for the five \name{} models of Section~\ref{sec:e-gen}. The storage figures count the expert-private indices and coefficient matrices, the shared basis and code book, and padding and alignment. All models use $K{=}256$ with \texttt{uint8} indices and the $(d,r)$ of Table~\ref{tab:eval-setup}, which shrinks the gate and up weights by $3.92$--$4.92\times$ on five models. Closed-form initialization recovers most of the quality lost to the coarse VQ. 
Short end-to-end distillation then restores PPL to the BF16-teacher level across all five models.
On four zero-shot downstream tasks, \name{} limits
Qwen3-30B-A3B's accuracy loss to at most 3.2 percentage
points relative to original BF16 model 
(Appendix~\ref{app:downstream}).


\begin{table}[t]
\centering
\caption{Quality and compression of the deployed configurations ($K{=}256$).
``Init.'' is closed form; ``KD'' is knowledge distillation.}
\label{tab:quality-generalization}
\footnotesize
\setlength{\tabcolsep}{2.4pt}
\begin{tabular*}{\columnwidth}{@{\extracolsep{\fill}}lcccccc@{}}
\hline
model & $(d,r)$ & comp. & teacher & init. & KD & $\Delta$PPL \\
\hline
Qwen3-30B-A3B & $(8,384)$ & $3.94\times$ & 9.82 & 12.51 & \textbf{9.56} & $-2.7\%$ \\
DeepSeek-V2-Lite & $(8,384)$ & $3.93\times$ & 8.37 & 12.49 & \textbf{8.22} & $-1.8\%$ \\
Nemotron-3-Nano & $(8,512)$ & $3.92\times$ & 8.55 & 9.66 & \textbf{8.29} & $-3.0\%$ \\
OLMoE-1B-7B & $(4,256)$ & $3.94\times$ & 8.33 & 9.11 & \textbf{8.24} & $-1.1\%$ \\
Gemma-4-26B-A4B & $(8,384)$ & $4.92\times$ & 10.98 & 13.37 & \textbf{10.44} & $-4.9\%$ \\
\hline
\end{tabular*}
\end{table}

\phantomsection\label{sec:e-correctness}
\noindent\textbf{Kernel fidelity.} 
We separately verify the numerical fidelity of the Trainium implementation
against an FP32 PyTorch reference.
For each model, we execute one MoE layer on Trainium using the fitted
$\mathcal B$, $J_e$, $D$, and $C_e$ and the model's actual dimensions,
including the gate, up, and down projections and the weighted sum over
the activated experts.
Compared with the same computation in FP32 PyTorch, the normalized maximum
error, 
$\max|\hat{y}-y|/\max|y|$, is
$3.40\times10^{-3}$ for Qwen3-30B-A3B,
$4.70\times10^{-3}$ for DeepSeek-V2-Lite,
$2.36\times10^{-3}$ for Nemotron-3-Nano-30B-A3B,
$4.90\times10^{-3}$ for OLMoE-1B-7B,
and $2.79\times10^{-3}$ for Gemma-4-26B-A4B.
The Nemotron-3-Nano-30B-A3B test covers its up-only ReLU$^2$ experts, whose hidden dimension is padded for the VQ blocks.

\phantomsection\label{sec:e-eq}

Table~\ref{tab:quality-attribution} compares stronger VQ-only configurations with $d=4$ and $d=2$ against the deployed $d=8$ ablation with the LR removed.  VQ-only with $d=2$ reaches PPL 10.44, 6.3\% above the BF16 teacher. 


\begin{table}[t]
\centering
\caption{Qwen3-30B-A3B quality attribution.. For \name, $(8,384)$ denotes $(d,r)$. VQ-only $d{=}8$ removes the LR component; smaller $d$ increases storage. Indexed-selection counts are relative to $d{=}8$.}
\vspace{-5pt}
\label{tab:quality-attribution}
\scriptsize
\setlength{\tabcolsep}{1.0pt}
\begin{tabular}{@{}lccccr@{}}
\hline
representation & training & \shortstack{bits/\\weight} &
\shortstack{indexed\\selections} & PPL & $\Delta$PPL \\
\hline
BF16 teacher & --- & 16.0 & --- & 9.82 & --- \\
shared LR ($r{=}512$) & closed form & 4.08 & --- & 34.91 & $+255.4\%$ \\
VQ only ($d{=}8$) & none & 1.00 & $1\times$ & 451.53 & $+4496.3\%$ \\
VQ only ($d{=}4$) & none & 2.00 & $2\times$ & 22.47 & $+128.7\%$ \\
VQ only ($d{=}2$) & none & 4.00 & $4\times$ & 10.44 & $+6.3\%$ \\
\name{} $(8,384)$ & closed form & 4.06 & $1\times$ & 12.51 & $+27.3\%$ \\
\textbf{\name{} $(8,384)$} & distillation & 4.06 & $1\times$ & \textbf{9.56} & $\mathbf{-2.7\%}$ \\
\hline
\end{tabular}
\vspace{-10pt}
\end{table}

\section{Related Work }

\label{sec:related}

\noindent\textbf{Scratchpad accelerators and workload placement.} From systolic arrays and flexible-dataflow designs~\cite{kung1982systolic,jouppi2017tpu,chen2016eyeriss,kwon2018maeri,qin2020sigma} to modern commercial systems (TPU, Trainium, Ascend, Tenstorrent, Gaudi)~\cite{aws2026trainium3arch,jax2026tpupipelining,huawei2026ascendcarchitecture,tenstorrent2026metalium,intel2022gaudi2}, accelerators couple dense compute with software-managed on-chip memory and heterogeneous engines, but generally take the stored tensor representation as given. Our question is complementary: how an MoE expert format changes the bytes and operations assigned to concurrent engines. LLM decode is memory-bound under the roofline due to low weight reuse~\cite{williams2009roofline,pope2023inference}, and overlap determines which work reaches latency~\cite{pati2024t3}. Other systems change the memory path: SN40L streams composition-of-experts models through tiered memory~\cite{prabhakar2024sn40l}, NPU--PIM partitions bandwidth-heavy operators~\cite{heo2024neupims,yun2024duplex}, and near-data MoE moves activations to cold experts~\cite{kim2024monde}. With the commercial accelerator fixed, \name instead co-designs the expert representation and decode dataflow around DMA, tensor-engine work, and indexed-selection work.

\noindent\textbf{Low-rank MoE factorization.} Activation-aware SVD~\cite{svdllm2025}, shared-basis MoE factorization~\cite{molae2025,chen2025mobe}, and architectural shared experts~\cite{dai2024deepseekmoe} reuse dense factors across experts. Shared factors are regular dense contractions that cut expert-private traffic, but confining every expert to one subspace caps capacity---the quality wall evaluated in Section~\ref{sec:e-kd}. \name retains this dataflow only to correct the VQ component's output error, not to approximate the whole weight.

\noindent\textbf{Weight vector quantization.} Scalar post-training quantization~\cite{frantar2023gptq}, product quantization~\cite{jegou2011pq}, and multi-code book, lattice, or trellis LLM quantizers~\cite{egiazarian2024aqlm,tseng2024quip,tseng2024qtip,vanbaalen2024gptvq,liu2024vptq} preserve high-rank structure while optimizing accuracy at a storage rate for GPU decode. RQ-MoE builds input-dependent code books for compressed embeddings~\cite{rqmoe2026}; \name compresses static expert weights with expert-private index maps. Low-rank quantization-error corrections fit factors per matrix~\cite{yao2023zeroquantv2,zhang2024lqer,saha2024caldera,zhang2025qera}, with LQER motivating dense correction over irregular high-precision gathers. We pool post-VQ residuals into one shared basis and, unlike bits-only comparisons, price indexed selection against DMA overlap: finer code vectors save HBM traffic but can make selection the bottleneck.

\noindent\textbf{MoE serving and distillation.} QMoE co-designs a sub-bit dictionary with GPU decode kernels~\cite{frantar2023qmoe}; DeepSpeed-MoE and MegaBlocks optimize placement, communication, and routed GEMMs~\cite{rajbhandari2022deepspeedmoe,gale2023megablocks}; and pre-gated MoE hides expert offload through prefetching~\cite{hwang2024pregated}. \name instead decides which representation state is shared, routed, and decoded by each engine. Knowledge distillation~\cite{hinton2015distilling} recovers the final quality gap but is not our contribution.
\section{Conclusion}
\label{sec:concl}

On an STA, an MoE expert representation defines its hardware dataflow. \name co-designs the representation with decoder execution to break the dependency chain from routing through weight transfer to computation: layer-shared factors expose routing independent work, while expert-private indices and coefficients preserve expert-specific capacity on the routing dependent path. This separation allows shared-factor transfers to overlap routing, shared projections to overlap expert-private transfers, and regular and indexed work to execute concurrently on the tensor engine and GpSimd. Across five MoE families, this co-designed dataflow delivers $1.15$--$1.31\times$ decode speedup with PPL at or below that of the BF16 teachers. More broadly, \name shows that model representation and STA dataflow should be designed together around the critical path, data movement, scratchpad residency, heterogeneous engine work, and synchronization.

\clearpage
\bibliographystyle{ACM-Reference-Format}
\bibliography{refs}
\clearpage
\appendix
\section{Appendix}
\label{app:details}

  \begin{figure}[t]
    \centering
    \includegraphics[width=\columnwidth]{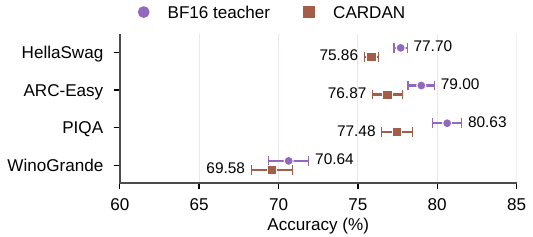}
    \caption{Zero-shot accuracy of Qwen3-30B-A3B.
    HellaSwag, ARC-Easy, and PIQA use length-normalized accuracy;
    WinoGrande uses unnormalized accuracy.}
    \Description{Paired points with horizontal error bars compare BF16-teacher and CARDAN zero-shot accuracy on HellaSwag, ARC-Easy, PIQA, and WinoGrande. The BF16 teacher has higher accuracy on all four plotted tasks.}
  \label{fig:downstream-accuracy}
  \end{figure}

\subsection{Whitened Shared-Basis Correction}
\label{app:fit}
Single-matrix low-rank quantization-error correction is
established~\cite{yao2023zeroquantv2,saha2024caldera}, including an
activation-weighted closed form in QERA~\cite{zhang2025qera}.
Equation~\eqref{eq:basis} instead fits one subspace to the pooled expert Gram,
yielding a layer-shared $D$.

For completeness, define the post-VQ residual
$\Delta_e=W_e-Q_{\mathcal B}(J_e)$. The shared-basis correction $(D,\{C_e\})$ minimizes
the activation-weighted output error
\begin{equation}
\label{eq:objective}
  \min_{D,\{C_e\}}
  \sum_e w_e\left\|\Sigma_x^{1/2}(\Delta_e-DC_e)\right\|_F^2,
  \qquad \Sigma_x=\mathbb E[hh^{\top}].
\end{equation}
Let $R=\operatorname{chol}(\Sigma_x)$, with $\Sigma_x=RR^T$. The route-aware
variant of Eq.~\eqref{eq:objective} uses
$w_e=\mathbb E[\mathbf 1(e\in\mathcal S(h))a_e^2]$; the default sets all
$w_e=1$. Let
\begin{equation}
\label{eq:basis}
 U_r=\operatorname{top\text{-}r}
 \left(R^T\left(\sum_e w_e\Delta_e\Delta_e^T\right)R\right).
\end{equation}
The shared basis and expert coefficient matrices are
\begin{equation}
\label{eq:codes}
 D=R^{-T}U_r,\qquad
 C_e=(D^T\Sigma_xD)^{-1}D^T\Sigma_x\Delta_e.
\end{equation}

To see optimality, whiten each post-VQ residual as $A_e=R^T\Delta_e$. For fixed VQ
factors, Eq.~\eqref{eq:objective} becomes a weighted shared-subspace
approximation of the matrices $A_e$. The Ky Fan/Eckart--Young result selects the
top-$r$ eigenspace of their pooled weighted Gram, giving Eq.~\eqref{eq:basis};
ordinary least squares then gives Eq.~\eqref{eq:codes}. This proves optimality
for the calibration objective under a common activation covariance. It does not
imply monotonic validation PPL.

\subsection{Exact rate accounting}
\label{app:storage}
The total storage rate per original weight, including the
expert-private $J_e$ and $C_e$ and the layer-shared
$\mathcal B$ and $D$, is
\begin{equation}
\label{eq:bits-full}
 b_{\mathrm{total}}=
 \frac{\log_2K}{d}+\frac{b_Cr}{H}
 +\frac{b_{\mathcal B}Kd+b_DHr}{EHI}.
\end{equation}
The final term is amortized across $E$ experts but is included in every reported model-size number. The implementation uses one-byte indices for $K=256$ and BF16 for deployed $\mathcal B$, $C_e$, and $D$; training checkpoints retain FP32 master copies that are not counted as inference storage.

\subsection{Exact Engine-Work Counts}
\label{app:work-vector}
Table~\ref{tab:work-vector-counts} summarizes how the representation
parameters affect the leading-order decode work on each engine for one projection. Shared counts apply once per token, while expert-private counts apply to each selected expert. Let $b_J(K)$ denote the deployed index width; $b_J(256)=8$ in our implementation, and the width changes only when $K$ crosses an encoding boundary. For compactness, Table ~\ref{tab:work-vector-counts} uses PE to denote the tensor engine.

\begin{table}[!ht]
\centering
\captionsetup{skip=4pt}
\caption{Representation parameters and leading-order decode work for one projection.}
\label{tab:work-vector-counts}
\footnotesize
\setlength{\tabcolsep}{2pt}
\renewcommand{\arraystretch}{1.05}
\begin{tabular*}{\columnwidth}{@{\extracolsep{\fill}}llll@{}}
\hline
scope / engine & operation & count & trend \\
\hline
shared PE & code book projection & $HK$ MACs & $K\!\uparrow\Rightarrow\uparrow$ \\
shared PE & basis projection $D^\top h$ & $Hr$ MACs & $r\!\uparrow\Rightarrow\uparrow$ \\
shared DMA & load $\mathcal B$ & $16Kd$ bits & $d\!\uparrow\Rightarrow\uparrow,\ K\!\uparrow\Rightarrow\uparrow$ \\
shared DMA & load $D$ & $16Hr$ bits & $r\!\uparrow\Rightarrow\uparrow$ \\
\shortstack[l]{expert-private\\DMA} & load $J_e$ & $\frac{HI}{d}b_J(K)$ bits & $d\!\uparrow\Rightarrow\downarrow,\ b_J\!\uparrow\Rightarrow\uparrow$ \\
\shortstack[l]{expert-private\\DMA} & load $C_e$ & $16Ir$ bits & $r\!\uparrow\Rightarrow\uparrow$ \\
\shortstack[l]{expert-private\\GpSimd} & indexed selection & $\frac{HI}{d}$ selections & $d\!\uparrow\Rightarrow\downarrow$ \\
\shortstack[l]{expert-private\\PE} & VQ reduction & $\frac{HI}{d}$ adds & $d\!\uparrow\Rightarrow\downarrow$ \\
\shortstack[l]{expert-private\\PE} & low-rank $C_e^\top z$ & $Ir$ MACs & $r\!\uparrow\Rightarrow\uparrow$ \\
\hline
\end{tabular*}
\end{table}

\subsection{Native MX at the BF16 Down Boundary}
\label{app:mx-boundary}

Our native selective Qwen W4A8 control preserves top-8 routing at TP${=}4$ and
LNC${=}2$: gate/up weights use MXFP4, their shared BF16 input is quantized online to
MXFP8 once per layer, and down remains BF16.  The W4A8 control therefore targets the same
routed gate/up traffic as \name{} without evaluating all experts or changing
the scheduler.

The MX primitive emits an accelerator-specific x4-packed representation (the
4P+Q layout in our implementation), which can remain in SBUF for an MX
consumer.  BF16 down instead requires an on-chip conversion of every selected
expert's output; this is not an HBM materialization.  In the compiled six-layer
control, replacing the MX down consumer with BF16 increases
\textsc{StreamTranspose} operations from 1 to 33 and \textsc{TensorCopy} from
34 to 110 per logical kernel.  We report this as a layout-boundary result, not
as evidence that MXFP4 arithmetic is intrinsically slow.


\subsection{Downstream Task Accuracy}
\label{app:downstream}
Using version 0.4.13 of the Language Model Evaluation Harness, we evaluate Qwen3-30B-A3B on HellaSwag, ARC-Easy, PIQA, and WinoGrande-XL in a zero-shot setting.
The BF16 teacher and \name{} checkpoint $(d,r)=(8,384)$ share the same tokenizer, request ordering. Figure~\ref{fig:downstream-accuracy} shows the comparison.

\end{document}